# Fabrication of magnetic tunnel junctions connected through a continuous free layer to enable spin logic devices

Danny Wan[1*], Mauricio Manfrini[1], Adrien Vaysset[1], Laurent Souriau[1], Lennaert Wouters[1], Arame Thiam[1], Eline Raymenants[1,2], Safak Sayan[1], Julien Jussot[1], Johan Swerts[1], Sebastien Couet[1], Nouredine Rassoul[1], Khashayar Babaei Gavan[1], Kristof Paredis[1], Cedric Huyghebaert[1], Monique Ercken[1], Christopher J. Wilson[1], Dan Mocuta[1], and Iuliana P. Radu[1]

[1]*Imec, Kapeldreef 75, Leuven, 3001, Belgium*

[2]*KU Leuven, Kasteelpark Arenberg, Leuven, 3001, Belgium*

[*]E-mail: danny.wan@imec.be

Magnetic tunnel junctions (MTJs) interconnected via a continuous ferromagnetic free layer were fabricated for Spin Torque Majority Gate (STMG) logic. The MTJs are biased independently and show magnetoelectric response under spin transfer torque. The electrical control of these devices paves the way to future spin logic devices based on domain wall (DW) motion. In particular, it is a significant step towards the realization of a majority gate. To our knowledge, this is the first fabrication of a cross-shaped free layer shared by several perpendicular MTJs. The fabrication process can be generalized to any geometry and any number of MTJs. Thus, this framework can be applied to other spin logic concepts based on magnetic interconnect. Moreover, it allows exploration of spin dynamics for logic applications.

# 1. Introduction

Advances of integrated circuits technology has been following Moore's law for about four decades.[1] However, downscaling is increasingly challenging as it will soon reach fundamental limits.[2-3] This has sparked active research in the field of "beyond-CMOS," seeking energy-efficient and scalable devices and architectures. Among these solutions, spintronics holds promise for low power manipulation of spin with current[4-6] and voltage.[7] Its inherent non-volatility can be used for high-density data storage,[8-9] but could also prove to be disruptive when combined with logic.[10-18] In particular, Spin Torque Majority Gate (STMG)[19] is a spin logic concept that could be employed for advanced logic circuits. Majority computation allows for greater circuit compactness[20] and could massively simplify device routing configurations. By cascading STMGs through a continuous ferromagnetic free layer (FL), information could propagate at high speed and low-power. Micromagnetic studies have explored this concept in detail to determine optimal shape[21] and operating window of this new device.[22-23] However experimental realization has remained elusive.

The prototypical STMG device consists of a cross-shaped FL connected to four independently addressable MTJs (Fig. 1a). STMG operation is as follows: magnetization in the FL is written via STT (Spin-Transfer Torque) generated at three of the MTJs acting as inputs. Below the MTJs, domain walls nucleate, merge, and propagate laterally to the fourth MTJ, which acts as an output and is sensed via tunneling magnetoresistance (TMR). STMG uses the same materials as conventional MRAM technology and is entirely compatible with CMOS (Complementary Metal-Oxide-Semiconductor) integration, making it a promising candidate for future logic technologies. These devices are interesting from a fundamental standpoint because they can be configured to study local magnetization switching in an extended free layer and provide a route to manipulate domain walls down to nanometer length-scales. With this research, we can investigate the underlying physics of domain wall dynamics and advance the field of spin based logic devices.

In this paper, we report on the fabrication of MTJs connected through a continuous FL geometry based on the STMG device architecture. Through this process, fully integrated STMG cross devices have been realized. We discuss some of the fabrication challenges and present preliminary measurements and initial device characterization. Finally, we end with a discussion on device scaling and show micromagnetic simulations that suggest majority gate operation is most amenable for devices with scaled dimensions.

## 2. Experimental methods

The fabrication process is detailed in Figure 1c and was presented at the 2017 International Conference on Solid State Devices and Materials.[24] Devices in this work were fabricated on imec's state-of-the-art 300-mm pilot line. First, the W bottom electrode contact (BEC) was patterned using a conventional MRAM (Magnetoresistive Random Access Memory) BEC process onto thermally oxidized Si substrates. The BEC was then capped by an ultra-smooth TaN seed layer to facilitate MTJ stack deposition. MTJs were deposited using an UHV (Ultra High Vacuum) multitarget magnetron sputtering system at room temperature without breaking vacuum between depositions. Stacking structure consists of perpendicularly magnetized dual-MgO MTJs, that are top-pinned by a reference layer (RL) and a synthetic ferromagnetic (SAF) system. The FL and RL are CoFeB-based and the SAF is composed of Co/Pt multilayers. The deposited MTJ films were measured with current in-plane tunneling technique resulting in TMR of 110% and resistance-area product of approximately 10 $\Omega \cdot \mu m^2$ (not shown). Wafers were then annealed at 300°C for 30 min in a 1 T perpendicular field.

STMG cross devices were patterned into the MTJ stack in two steps. First, pillars were defined using 193nm immersion (193i) lithography combined with Ar ion beam etching at normal and grazing angle.[25] The pillar patterning step consists of a challenging etch of the magnetic stack stopping on MgO tunnel barrier to electrically isolate neighboring MTJs from each other. Next, 193i double patterning lithography and ion beam etching were used to define cross-shaped FLs and to magnetically isolate devices. Finally, top electrode contacts were made for electrical access to devices using a modified BEOL (Back End of Line) flow compatible with MTJ thermal budgets.[26] Finally, trenches and vias were patterned by employing a dual damascene etch followed by standard Cu metallization. After fabrication, wafers were submitted for magnetic reset in a 2 T perpendicular field at room temperature.

Figure 2a shows a high resolution cross-sectional TEM (Transmission Electron Microscopy) image of a completed device with the cross structure bisected through the long axis. MTJs are shown contacted above by independent top electrodes and connected below through a damage-free, continuous FL.

## 3. Results and discussion

### 3.1 MTJ pillar patterning details and peak force tunneling AFM of etched cross-sections

In Figure 2b, a thin layer of residual RL material is present above the MgO, forming a bridge between the base of MTJs. This residual RL material was intentionally left as a buffer layer during pillar definition. Ion beam etching employs high energy Ar ions that can penetrate up to 3 nm into the stack and induce crystallinity damage of MgO for the chosen energy. Therefore, instead of stopping directly on the barrier, we preemptively terminate the process, leaving a few nm of unetched RL stack. In doing so, we avoid ion bombardment of the film and preserve interface-induced PMA.[27-28]

This residual RL can act as a parasitic shorting pathway between neighboring MTJs, which impacts TMR, diminishes STT, and influences DW propagation. To address this issue, we applied post etch plasma oxidation to convert the residual RL material into electrically insulating and magnetically inactive films. Successful conversion was verified using peak-force tunneling atomic force microscopy (PF-TUNA)[29] to measure the MTJ cross-section on blanket wafers before etching and after etching with oxidation. Two-dimensional current maps were obtained showing the presence of RL and SAF materials before etching (Fig. 2c). In contrast, this residual material is no longer conductive in the post etch and oxidation maps, which confirms that the oxidation step successfully converted metallic layers sitting on top of the MgO into insulating material.

### 3.2 FL patterning details and presence of sidewall fencing

After patterning MTJ pillars, the FL is patterned to isolate individual cross structures. It is known that FL geometry can play a significant role in magnetization dynamics where a non-ideal FL shape can suppress switching.[21] The FL target critical dimension is 50 nm, which is approaching the limitations of single print 193i lithography. To bypass this limitation, we have employed double patterning to define each arm of the cross independently (Fig. 3a). The resulting pattern is then transferred into spin-on-carbon (SoC) as an intermediate hard mask material (Fig. 3b), which is subsequently used to transfer the cross shape through the FL, stopping in W BEC. During ion beam etching, stack materials are unavoidably resputtered, and produce a prominent sidewall fence (Fig. 3c) which cannot be fully removed.

XTEM (Cross-Sectional Transmission Electron Microscopy) (Fig. 3d) and EDS (Energy-dispersive X-ray Spectroscopy) chemical mapping (Fig. 3e) confirm that the composition of the fence is W originating from the BEC. This metallic sidewall residue will create a parasitic short across the tunnel barrier, lowering TMR and impacting device characteristics. Attempts have been made to remove or diminish the fencing, however a shorting path remains. Optimizing the etch process and limiting overetch could reduce W redeposition. Replacing SoC with $SiO_2$ as the hardmask material could also improve the fencing problem. $SiO_2$ is a more robust material, which would make it more resistant than SoC to penetration of resputtered material.

### 3.3 Magneto-electrical characterization

Fully integrated devices on 300-mm wafers have been demonstrated. Cross-shaped devices have been measured addressing each MTJ independently. Figure 4 shows the magneto-electrical characterization performed on our devices. The device under test is comprised of four pillars of 70 nm diameter, separated center-to-center by 300 nm. The edge of the FL is located 50 nm further than the edge of the pillar, as shown in Figure 4a. Although it is desired that all pillars within the same device present similar TMR values, in our devices there is a strong variation. One possible explanation for the TMR values would be the two-step FL patterning that first defines the FL for pillar 1 (P1) and pillar 4 (P4), followed by the definition of the FL for pillar 2 (P2) and pillar 3 (P3), as shown in Figure 3a. However, Figure 4b shows that TMR values do not follow a pattern, with TMR of P3 and P4 larger than TMR values of P1 and P2. Another scenario for the observed asymmetric TMR values can be due to the fencing generated during FL etching. Stray materials not only cause device shorting through parasitic pathways, but also influence the orientation of magnetic moments in the core of the pillar, resulting in decreased TMR.[30] Since the uniformity of the fencing around the free layer cannot be guaranteed, TMR values in each pillar can be individually different. Moreover, one can note that the FL coercive field (Hc) values for each pillar are distinct, giving a clear indication that the magnetization reversal in the FL below the pillar follows different dynamics. This is evidence of structural defects in the FL, hindering DW propagation in the device.

To verify that all pillars within the cross geometry can be STT functional, we studied the STT excitation in individual pillars. Each pillar is biased with 1 $\mu$s long pulses, and the voltage range is swept. Resistance versus voltage curves for each pillar, normalized to TMR values, are depicted in Figure 4c. The curves have been normalized to TMR values to improve visualization. Device resistances are shown in Figure 4d. Despite low TMR values at device level, it is possible to observe the parallel to antiparallel (P2AP) transition for all four pillars. This is an important result that indicates that electrically-controlled magnetization reversal at the free layer below each pillar was achieved. It is also a signature of DWs being nucleated with MTJs via STT, a crucial component towards the realization of STMG technology. For the antiparallel to parallel (AP2P) transition, the magnetization dynamics is a bit different. Only P1 and P4 reverse to parallel state, whereas P2 and P3 experiences MgO breakdown process, given the gradual resistance reduction. It is imperative that in future devices, the switching voltages are reduced to not overlap with the MgO breakdown values (0.8 V to 1.3 V). Note also that for P4, when the voltage is near to the switching voltage threshold, the device resistance increases prior to switching. One possible explanation is that the DW sitting below P4 is not profiting from full TMR and it moves to cover the entire region below the pillar prior to switching. To further understand the discrepancies in TMR values and STT switching behavior, we study at device level the RA product dependence on the bias voltage, as depicted in Figure 4d. Compared to blanket RA value (10 $\Omega.\mu m^2$), the devices show a reduced RA product, most likely due to sidewall short paths. Particularly, P2 is heavily shorted. However, P3 shows a high resistance after P2AP transition but it does not transition back to parallel state for the applied voltage range. This shows that the fencing around the devices is inhomogeneous, yielding different switching behaviors for the MTJs. Therefore, real-time majority operation with DWs remains elusive due to undesired processing conditions of our devices. Also, the current device dimensions have greater tendency for STMG operation to fail due in DW pinning, which is exacerbated by larger cross area. Nevertheless, to show electrical device functionality for a system of four interconnected MTJs is a major step for STMG realization.

**3.4 Micromagnetic simulations towards ultra-scaled feature size**

Micromagnetic simulations show that STMG is expected to be functional if the lateral dimension $L$ is smaller than five times the DW width $\delta$,[23] defined as the critical size. This is due to the exchange-driven character of domain expansion since no lateral current is injected to drive DW.[22] Above this critical size $5\delta$, the nucleated DW is prone to pinning due to material defectivity and line edge roughness.[31-32] Below the critical size, the DW becomes unstable: magnetization is non-uniform during switching but converges to a uniform state after switching off the current. To reach the necessary critical dimensions, MTJ and FL patterning was also demonstrated using electron beam lithography (EBL). Patterning fidelity and line roughness using EBL improved significantly compared to the immersion process and critical dimension of less than 20 nm were achieved (Fig. 5a).

Using the OOMMF (Object Oriented MicroMagnetic Framework) package,[33] micromagnetic simulations have been carried out to study the impact of roughness and shape variability for scaled dimensions. Figure 5 shows a simulation for an arm width of 10 nm. The exchange constant is $A_{ex} = 2\times10^{-11}$ A/m and the effective anisotropy is $K_{eff} = 30$ kJ/m$^3$, which results in a DW width $\delta = \sqrt{A_{ex}/K_{eff}} = 26$ nm. The effective anisotropy is lower than for a typical STT-MRAM stack since the STMG stack is, in principle, optimized for speed rather than long-term data retention. In practice, such a low effective anisotropy value can be reached by tuning the free layer thickness. The lateral size $L = 90$ nm is smaller than the critical size $5\delta = 130$ nm and therefore fulfills the condition for a functional majority gate. STT is applied to reverse the magnetization in the upper and left arms into the down state, while it keeps the magnetization up in the lower arm. After about 3ns, we observe that the down state, imposed by the majority of input MTJs has propagated to the output arm of the cross. In fact, after releasing the current, the entire cross relaxes to the down state. This simulation was repeated for each input combination and for other shape variations. The majority behavior was confirmed in every case. Consequently, for dimensions smaller than the critical size $5\delta$, roughness does not seem to affect magnetization dynamics. This is likely due to the DW being too wide to be pinned by roughness.

## 4. Conclusions

We have fabricated MTJs interconnected via a continuous ferromagnetic free layer and characterized them using conventional magneto-electronic techniques. Each pillar can be addressed independently and the devices exhibit STT functionality and exhibit DW nucleation, which is a major step towards the realization of STMG technology. To realize majority gate function, DW pinning must be suppressed, which can be attained through device scaling. Future work is in progress to pattern ultra-scaled device dimensions towards demonstration of majority gate and real-time observation of DW propagation. In this work, we have presented the fabrication of a cross geometry with three inputs and one output, but this fabrication process could be generalized to any geometry with any number of inputs and outputs. Thus, this framework can be applied to other spin logic concepts based on magnetic interconnect and allows exploration of spin dynamics for logic applications.

## Acknowledgments

This work was performed as part of the imec IIAP core CMOS. The authors gratefully acknowledge V. De Heyn for operation support and P. Favia for TEM images.

## Figure Captions

**Fig. 1.** (Color Online) (a) Schematic of the target structure: a cross-shaped FL connected to 4 top-pinned, independently addressable MTJs. (b) Photograph of a fully integrated 300-mm wafer. (c) The device fabrication process flow.

**Fig. 2.** (Color Online) (a) High resolution cross-sectional TEM image of a completed device showing MTJs contacted above by TE (top electrode) and connected below by continuous FL. (b) Close up XTEM showing residual RL materials. (c) and (d) Tunneling AFM (Atomic Force Microscopy) 2D current maps of an MTJ stack overlaid with representative line cut before and after the oxidation process. The oxidation step successfully converted metallic layers (SAF + RL) on top of the MgO layer into an insulating material.

**Fig. 3.** (Color Online) (a) SEM (Scanning Electron Microscope) image showing dual hard masks used for FL double patterning. (b) Angled SEM view of cross pattern transferred into SoC. (c) Angled SEM view of cross pattern transferred to FL with residual fencing. (d-e) Close up XTEM and EDS maps showing that the fence contains metallic W.

**Fig. 4.** (Color Online) (a) Top-view and cross-section schematics of a cross-shaped STMG device. (b) Resistance versus field curves for each individual pillar and its (c) corresponding STT excitation response. Note due to the stray fields originating from the RL/SAF system ($\mu_0 H \sim 15$ mT), the MTJs are encountered at antiparallel state at zero field. (d) RA product at device level as function of bias voltage.

**Fig. 5.** (Color Online) (a) A cross patterned at 20nm using ebeam lithography (inset: comparison using 193i). (b-d) Micromagnetic simulation for an arm width of 10 nm at 0ns, 1ns, and 3ns. Magnetization of left and upper arm are switched to down state and show majority propagation to the right output. The white dashed lines show the position of the MTJs. Inside these areas, a current density $J = 6 \times 10^{11}$ A/m² is applied, with current flowing vertically.

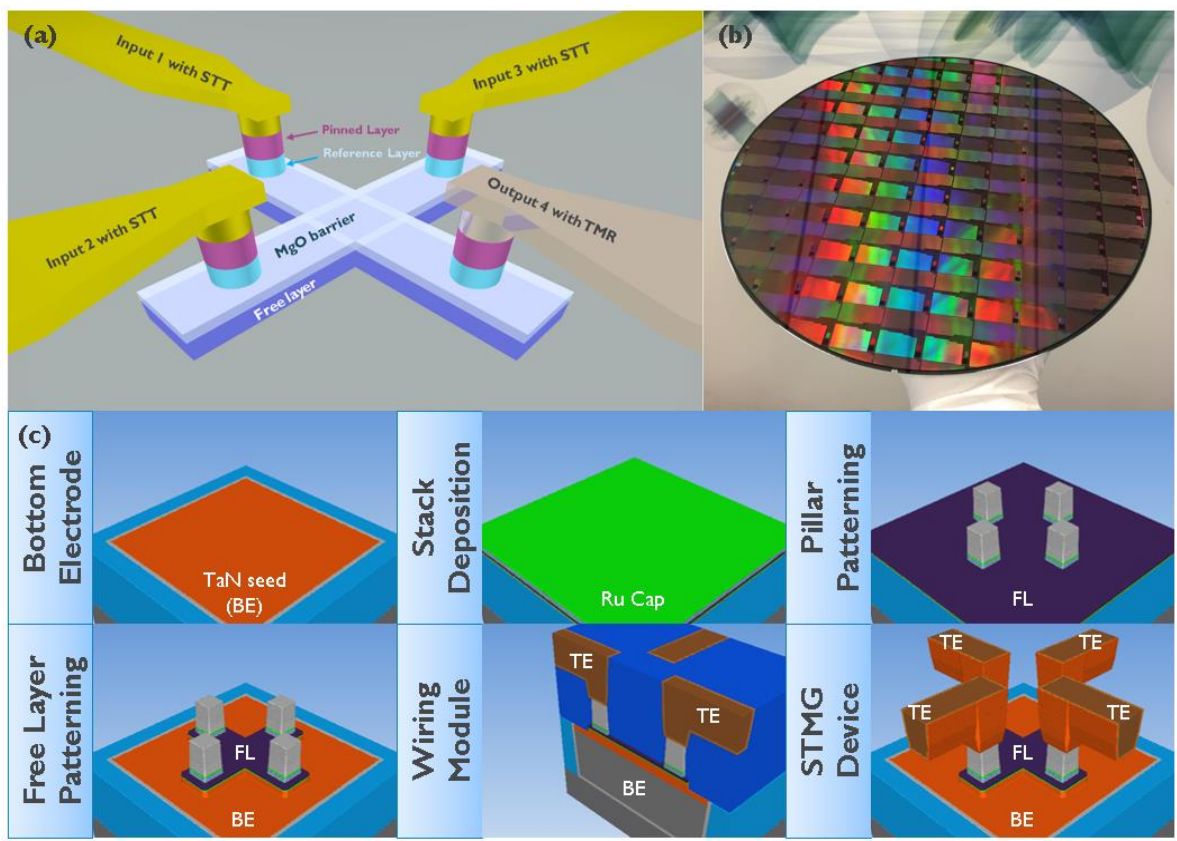

Fig.1. (Color Online)

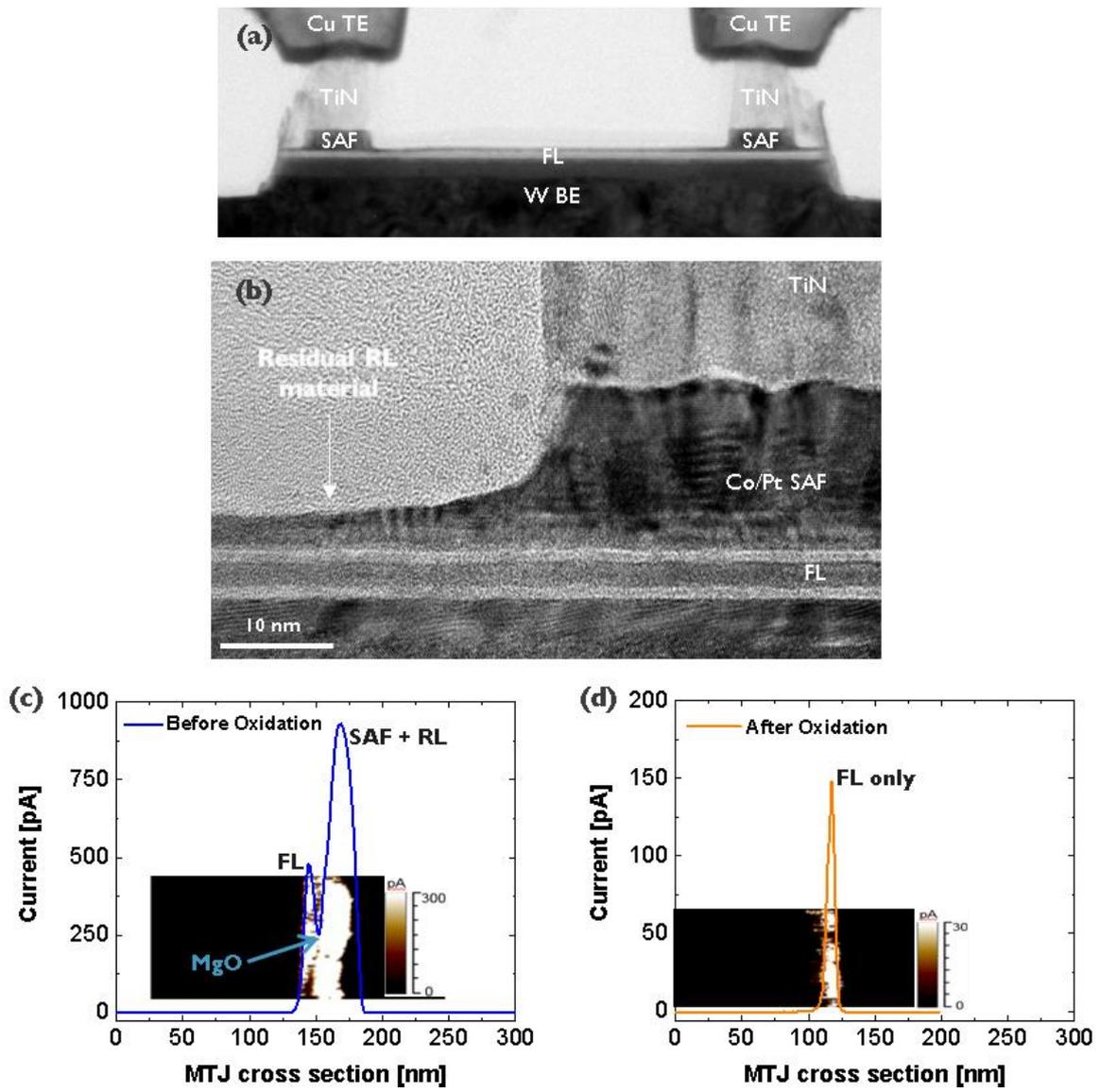

Fig.2. (Color Online)

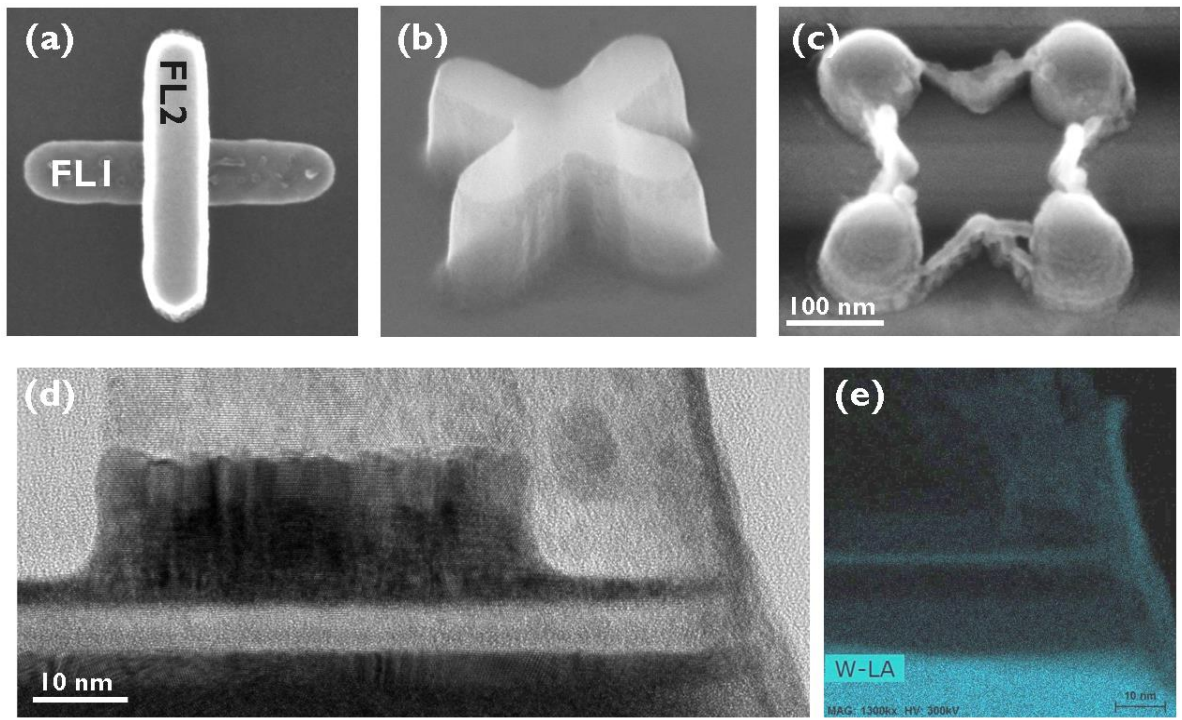

Fig.3. (Color Online)

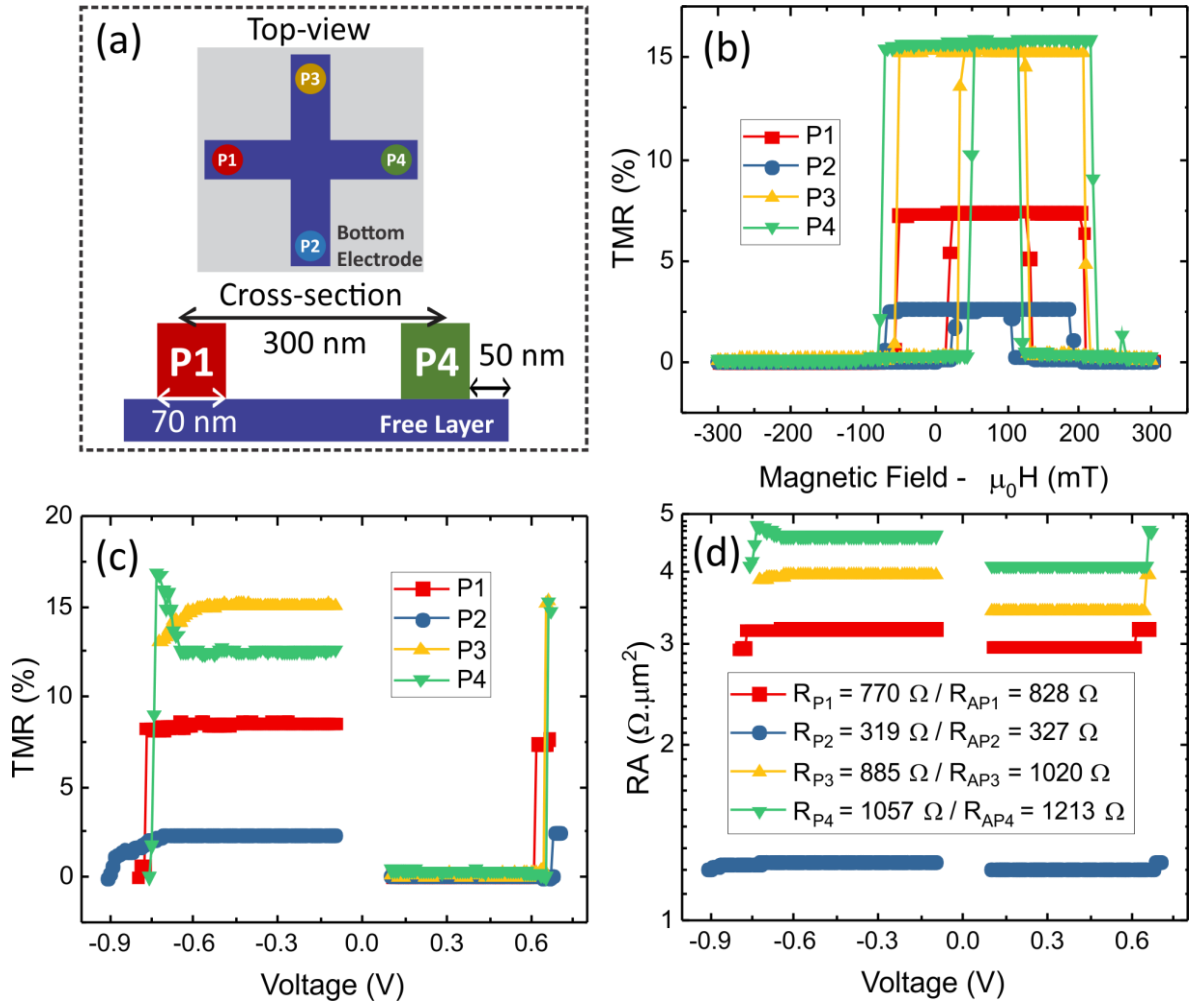

Fig.4. (Color Online)

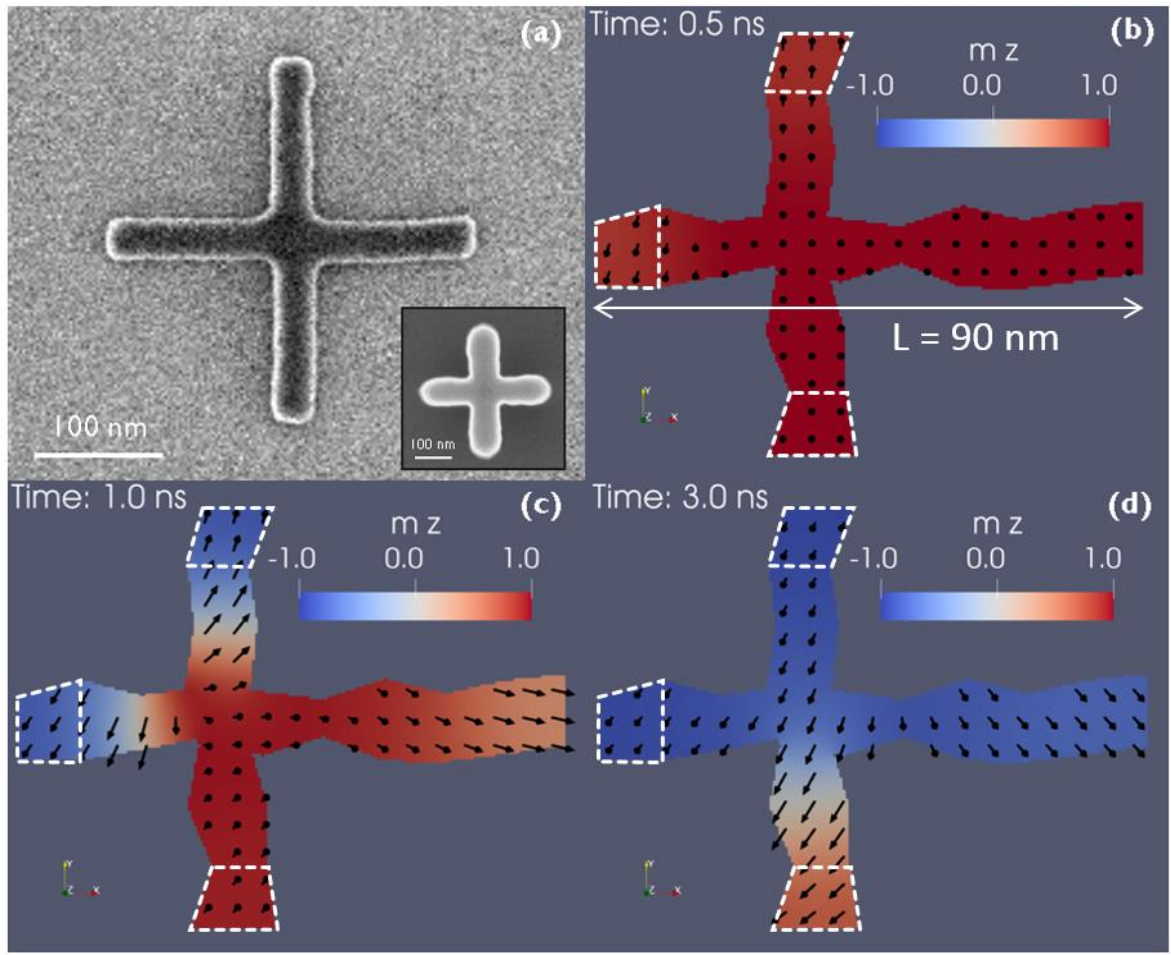

Fig.5. (Color Online)